\documentclass[preprint,rnote]{aa} 

\usepackage{txfonts} 
\usepackage{natbib} 
\usepackage{amstext}
\usepackage{graphicx}
\usepackage{array}

\newcommand{\reaction}[6]{\nuc{#1}{#2}(#3,#4)\/\nuc{#5}{#6}}

\newcommand{\nuc}[2]{\ensuremath{^{#1}}#2}
\newcommand{\Nepa}[0]{\ensuremath{^{22}}Ne\ensuremath{+ \alpha}}
\newcommand{\ag}[0]{\reaction{22}{Ne}{$\alpha$}{$\gamma$}{26}{Mg}}
\newcommand{\an}[0]{\reaction{22}{Ne}{$\alpha$}{n}{25}{Mg}}
\newcommand{\neon}[0]{\reaction{20}{Ne}{p}{$\gamma$}{21}{Na}}
\newcommand{\flou}[0]{\reaction{18}{F}{p}{$\alpha$}{15}{O}}
\newcommand{\magn}[0]{\reaction{25}{Mg}{p}{$\gamma$}{26}{Al}}

\begin{document}

\title{Recommendations for Monte Carlo nucleosynthesis sampling}
 
\author{R. Longland\inst{1,2}\thanks{\email{richard.longland@upc.edu}}}
       
\institute{
 Departament de F\'{\i}sica i Enginyeria Nuclear, EUETIB,
  Universitat Polit\`{e}cnica de Catalunya, c/~Comte d'Urgell 187,
  E-08036 Barcelona, Spain\label{inst1} \and
  Institut d'Estudis Espacials de Catalunya (IEEC), Ed. Nexus-201,
  C/~Gran Capit\`{a} 2-4, E-08034
  Barcelona, Spain\label{inst2}}

\abstract{
  Recent reaction rate evaluations include reaction rate uncertainties
  that have been determined in a statistically meaningful
  manner. Furthermore, reaction rate probability density distributions
  have been determined and published in the form of lognormal
  parameters with the specific goal of pursuing Monte Carlo
  nucleosynthesis studies. 
}{
  To test and assess different methods of randomly sampling over
  reaction rate probability densities and to determine the most
  accurate method for estimating elemental abundance uncertainties.
}{
  Experimental Monte Carlo reaction rates are first computed for the
  \Nepa, \neon, \magn, and \flou\ reactions, which are used to
  calculate reference nucleosynthesis yields for 16 nuclei affected by
  nucleosynthesis in massive stars and classical novae. Five different
  methods of randomly sampling over these reaction rate probability
  distributions are then developed, tested, and compared with the
  reference nucleosynthesis yields.  
}{
  Given that the reaction rate probability density distributions can
  be described accurately with a lognormal distribution, Monte Carlo
  nucleosynthesis variations arising from the parametrised estimates
  for the reaction rate variations agree remarkably well with those
  obtained from the true rate samples. Most significantly, the most
  simple parametrisation agrees within just a few percent,
  meaning that Monte Carlo nucleosynthesis studies can be performed
  reliably using lognormal parametrisations of reaction rate
  probability density functions.}{}


\maketitle

\section{Introduction}
\label{sec:intro}

With the growing availability of high power computers, our ability to
explore the effects of nuclear physics uncertainties on
nucleosynthesis has reached new levels. With post-processing
calculations, for example, thousands of computations can be performed
in relatively little time, thus opening new avenues for
exploration. Furthermore, better treatment of reaction rate
uncertainty propagation has recently lead to the publication of rates
that include statistically meaningful uncertainties
\citep{LON10,ILI10a,ILI10b,ILI10c} and include lognormal
representations of the rate probability density
functions\footnote{Note that there is no restriction that the rates
  obtained in \cite{ILI10a} must be within the published $1\sigma$
  ``high'' and ``low'' rates but rather follow a continuous
  probability density distribution.}. Given that information, methods
should be developed to utilise those reaction rate uncertainties and
investigate their effect on nucleosynthesis.

Monte Carlo methods are one option for investigating the effects of
reaction rate uncertainties on nucleosynthesis yields. These methods
can either be used to investigate the effects of a single reaction, a
small group of reactions, or they can be used to investigate the
overall effect of all reaction rates in complex environments involving
a great number of competing reactions. The general strategy in the
latter case is to (i) compute a sample rate for every uncertain
reaction simultaneously, (ii) compute nucleosynthesis yields using
these samples in a post-processing model, and (iii) repeat many
times. Correlations between abundances and reaction rates can be found
in a post-analysis procedure to identify reaction rates of
interest. These methods have be used previously in a number of studies
\citep[e.g.,][]{HIX03,ROB06,PAR08}. In each of those studies,
temperature independent enhancement factors were applied to the rates
of all nuclear processes to investigate their effect on
nucleosynthesis in novae and x-ray bursts. The enhancement factors
were sampled from lognormal distributions that were chosen to
represent the estimated uncertainty of the reaction rates at
temperatures most important in these environments. In some studies,
these estimations were arrived at by considering very general
characteristics of the reaction rates such as whether they are derived
purely from theoretical predictions or were measured with radioactive
beams~\citep[e.g.,][]{HIX03}. Until now, however, these methods have
not considered the rate uncertainties for each reaction separately.

The method of varying reaction rates by a constant factor assumes that
the true, unknown, reaction rate differs from the recommended rate by
a constant factor over all temperatures. Equivalently, this method
makes the assumption that rate uncertainties are
temperature-independent. However, in reality, the uncertainties of
individual reaction rates are, indeed, often temperature dependent if
they are based on experimental constraints, usually with larger
uncertainties associated with the rate at lower temperatures. This can
clearly be seen in the uncertainty bands presented by
\cite{ILI10c}. Furthermore, by considering the way that reaction rates
are computed, it is clear that the temperature dependence of a sample
reaction rate should not necessarily follow the same temperature
dependence as the recommended rate. For example, the rate could be
higher than the recommended rate at some temperatures, while being
lower at others. 
How important is this effect? \cite{THE00} discussed the ``race
against time'' effect that different temperature dependencies for the
\Nepa\ reaction rates might have on the s-process in massive
stars. Presumably, the complex interaction and competition of reaction
rates in a dynamically changing environment means that simple
enhancement of reaction rates does not fully explore the
nucleosynthesis variations arising from reaction rate
uncertainties. More advanced methods should, therefore, be
investigated to fully account for this effect in nucleosynthesis
uncertainty studies. The rate probability density distributions
published in \cite{ILI10a} make this investigation possible, while
eliminating the need for simplifying estimations of rate uncertainties
based on general reaction properties.


The obvious choice for correctly performing Monte Carlo
nucleosynthesis studies is to directly use Monte Carlo reaction rate
samples obtained separately from the \texttt{RatesMC} code
\citep{LON10}. In this case, individual samples of the nuclear physics
input are used, thus accounting for all possible behaviour of the
reaction rates.
However, this requires a considerable amount of effort using tools
that most computational astrophysicists do not have access to. The
purpose of this research note, therefore, is to investigate different
rate sampling schemes in comparison to the ideal case discussed above,
and present the best practice with which to utilise the reaction rate
uncertainties that are published in \cite{ILI10a}. In
Sec.~\ref{sec:sampling}, a number of possible sampling schemes are
discussed. Those schemes are tested with an example set of reaction
rates and discussed in Sec.~\ref{sec:results}. Conclusions are given
in Sec.~\ref{sec:conclusions}.

\section{Reaction rate sampling methods}
\label{sec:sampling}

The reactions evaluated by \cite{ILI10a} involved target nuclei with
masses in the A$=14 - 40$ range, whose cross sections are usually
dominated by resonant capture.  For reactions that involve a large
number of resonances, it is clear that a great number of possibilities
exist for how a sample reaction rate could vary with respect to its
recommended value. In light of this, the most consistent way to sample
the reaction rates is to sample the nuclear physics input separately
for each nucleosynthesis calculation. Since most computational
investigations do not have access to the data necessary to perform
this kind of sampling, this method is viewed as the ideal, yet
unrealistically complicated standard. 
For the purpose of this study, the \texttt{RatesMC} code used by
\cite{LON10} is used to compute this ideal case for comparison
alongside the estimations considered in this work. Throughout the
following discussion, rate variations obtained from samples of the
nuclear physics uncertainties are referred to as the ``optimum'' rate
samples.

In order to construct rate variation schemes, we must first evaluate
the expected behaviour of the reaction rates as a function of
temperature. A reaction rate per particle pair is computed by
\begin{equation}
  \label{eq:rates-reacrate}
  \langle\sigma v\rangle = \sqrt{\frac{8}{\pi \mu}}
  \frac{1}{(kT)^{3/2}}\int_{0}^{\infty}E\sigma(E)e^{-E/kT} dE
\end{equation}
where $\mu$ is the reduced mass of the reacting particles,
$\mu=M_{0}M_{1}/(M_0+M_1)$; $M_i$ denotes the masses of the particles;
$k$ is the Boltzmann constant; $E$ is the centre-of-mass energy of the
reacting particles; and $\sigma(E)$ is the reaction cross section at
energy, $E$. This cross section frequently consists of a collection of
resonances, so the reaction rate varies according to resonance
properties, such as resonance strength or energy. The convolution of
these resonance cross sections with the Boltzmann distribution in
Eq.~(\ref{eq:rates-reacrate}) will \textit{always} result in a
smoothly varying reaction rate with temperature. 

To construct a rate variation scheme, first consider the available
information from \cite{ILI10a}; information that is also included
directly in the \texttt{starlib} library \citep[see, for
example,][]{ILI11}. The tabulated rates include two parameters, $\mu$
and $\sigma$, to describe the rate probability density distribution at
each temperature \citep[see][]{LON10}:
\begin{equation}
  \label{eq:mu-sigma}
  \mu = \ln{x_{\text{med}}}, \qquad \sigma=\ln \sqrt{\frac{x_{\text{high}}}{x_{\text{low}}}},
\end{equation}
where $x_{\text{med}}$, $x_{\text{high}}$, and $x_{\text{low}}$ are
the recommended (median), high, and low reaction rates,
respectively. The variables $\mu$ and $\sigma$ here refer to the
\textit{lognormal} shape parameters, and are not to be confused with
the more commonly used Gaussian mean and standard deviation. A
reaction rate sample, $x(T)$, at a specific temperature, $T$, can
be represented by
\begin{equation}
  \label{eq:rate}
  x(T) = e^{\mu(T)} \cdot e^{p(T) \sigma(T)}
\end{equation}
where $p(T)$ is a random variable that is normally distributed (i.e.,
distributed according to a Gaussian distribution with an expectation
value of 0 and standard deviation of 1). The second component of
Eq.~(\ref{eq:rate}) is dubbed as the ``uncertainty factor'', and is a
temperature dependent quantity. Given this information, the challenge
becomes one of choosing values for $p(T)$, recalling that these values
must vary smoothly with temperature, and must be distributed normally
to enforce the lognormality of reaction rate probability densities.

The simplest parametrisation for $p(T)$ is determined by assuming that
it is independent of temperature, i.e., $p(T) = a$ where $a$ is
sampled from a normal distribution. This is dubbed the ``flat''
parametrisation in the following discussion, although it is important
to emphasise that for this parametrisation, the uncertainty factor in
Eq.~(\ref{eq:rate}) is still temperature dependent, and given by
$e^{a\, \sigma(T)}$. This was not the case in the Monte Carlo
variations of previous studies
\citep[e.g.,][]{PAR08}. 
Those studies applied an uncertainty factor to the rate regardless of
the uncertainty temperature dependence. Their method is equivalent to
using a constant value, $f$, for the uncertainty factor. In this case,
Eq.~(\ref{eq:rate}) would become $x(T) = f \,e^{\mu(T)}$. As mentioned
before, this is not representative of the rate uncertainty at all
temperatures for experimentally constrained reaction rates. It is,
however, a good approximation for reaction rates derived purely from
theory.

\begin{figure}
  \centering
  \includegraphics[width=0.5\textwidth]{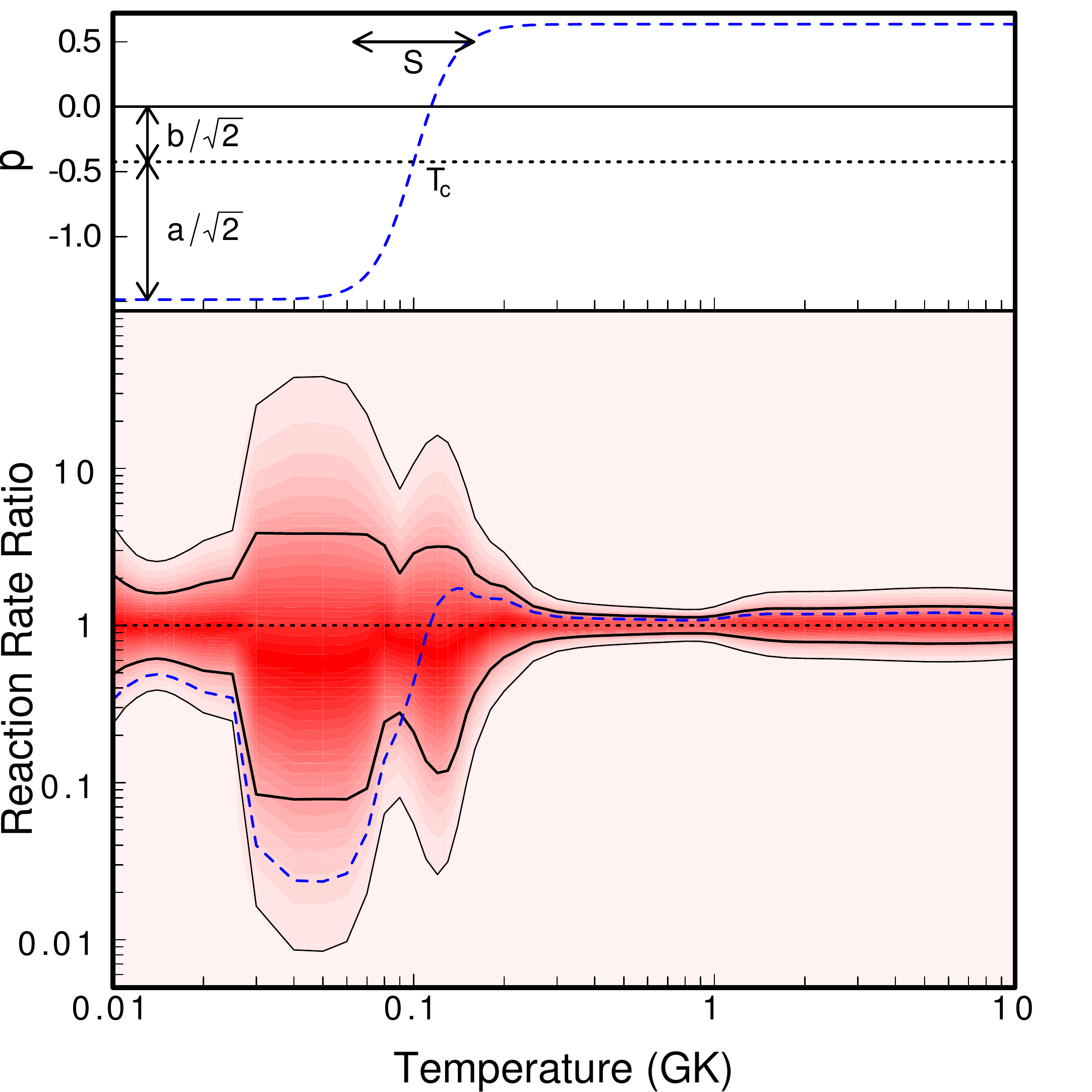}
  \caption{(Colour online) Example of a single rate sample of the \ag\
    reaction using the parametrised rate factor variations. Plotted is
    the reaction rate ratio (i.e., compared to the recommended
    rate). Therefore, the line at unity represents the recommended
    rate, while a point at 10 means that the rate is 10 times that of
    the recommended rate. The red density distribution represents the
    coverage probability of the reaction rate, with thick and thin
    black lines denoting the 68\% and 95\% uncertainties,
    respectively. The dashed blue line represents a single sample of
    the reaction rate obtained when using Eq.~(\ref{eq:rate}) with
    $p(T)$ determined from the parametrisation in Eq.~(\ref{eq:tanh})
    and shown in the inset.}
  \label{fig:Sample}
\end{figure}

Consider, now, the case in which the reaction rate uncertainty can be
characterised in two regions. A single rate sample could, for example,
be high with respect to the recommended rate at lower temperatures,
while being low at higher temperatures. This situation can be
reproduced by parameterising $p(T)$ in Eq. (\ref{eq:rate}) with a
hyperbolic tangent function. To further generalise this
parametrisation, an offset parameter can be utilised: 
\begin{equation}
  \label{eq:tanh}
  p(T) = \frac{1}{\sqrt{2}} \left(o + a \tanh \left[ \frac{2.94}{S}\log_{10}(T_{c}/T)\right]\right),
\end{equation}
where the sampled variables $a$, $o$, $S$, and $T_{c}$ are shown in
the inset of Fig.~\ref{fig:Sample} and are defined as the maximum
deviation, offset, spread parameter, and cross-over temperature,
respectively. The spread parameter is defined as the logarithm of the
temperature range needed for $p$ to vary from $p=o+0.9a$ to
$p=o-0.9a$. In this parametrisation scheme, $a$ and $o$ must be
sampled from normal probability density distributions, while $S$ and
$T_{c}$ can take on a range of values sampled from flat probability
density functions. Here, the ranges are chosen to be $0.1 < S < 1$~GK
and $10^{-2} < T_c < 10$~GK. This parametrisation is referred to as
the ``full'' parametrisation hereafter. An example of applying this
parametrisation to the \an\ reaction rate is shown in
Fig.~\ref{fig:Sample}, where $a=-1.5$, $o=-0.6$, $T_c=0.1$~GK, and
$S=0.4$~GK.


To further investigate the hyperbolic tangent parametrisation of the
reaction rate samples, two more cases are considered. In the first
case (the ``no offset'' case), the offset parameter is constrained to
zero. This requires removing the $1/\sqrt{2}$ factor in
Eq.~(\ref{eq:tanh}) to ensure normality of the distribution. The
parameter space for the reaction rate will be constrained for this
case, in particular, the possibility that the rate has a systematic
shift at all temperatures will not be accounted for. The
nucleosynthesis yield uncertainties determined from this method are
therefore expected to be reduced using this parametrisation. For the
second case (called ``fixed $S$ and $T_c$''), the parameters, $S$ and
$T_c$ are fixed at values of $S = 0.2$ and $T_c = 0.2$~GK. These
values are chosen to maximise the reaction rate shape variations
around $0.2$~GK. This final case should help answer the question of
whether the temperature dependence of the \Nepa\ reactions has a large
effect on s-process yields as predicted by \cite{THE00}.

Finally, although we have argued that $p(T)$ must vary smoothly with
temperature, it can be illustrative to investigate the effect on
nucleosynthesis when $p(T)$ this constraint is relaxed. On a grid of
temperatures, therefore, $p(T)$ is sampled independently so that the
resulting rate can fluctuate randomly from one temperature to the
next. These sample rates are obtained once per Monte Carlo
nucleosynthesis computation. 
This will be referred to as the ``random'' case. Another technique is
to sample all reaction rates at every evaluation step within the
nucleosynthesis code. In this case, the rate will change rapidly
around the recommended rate over small temperature ranges, thus
yielding unrealistically small nucleosynthesis uncertainties. This
latter case is not investigated here.

\section{Test Cases}
\label{sec:test-cases}

The effects of the 5 rate parametrisation schemes, in addition to the
optimum rate samples calculated directly from the nuclear physics input,
are investigated with respect to a test suite of reactions and
scenarios. To achieve this, post-processing calculations are used to
calculate Monte Carlo nucleosynthesis yields with some representative
profiles and networks. To obtain a good representation of the
uncertainties, 3000 Monte Carlo samples are used for each method,
which are found to reproduce abundances to within 3\% and
uncertainties to within about 10\% between runs.

The first of the test cases are the \Nepa\ reactions (i.e., \ag\ and
\an), which are important for s-process neutron production in massive
stars and asymptotic giant branch stars. For an overview of the
reactions and their nuclear physics uncertainties, the reader is
referred to, for example, \cite{JAE01,KOE02,LON12}. A
temperature-density profile corresponding to core helium burning in a
25~M$_{\odot}$ massive star is used to study nucleosynthesis for these
reactions, and has been discussed elsewhere by \cite{THE00} and
\cite{ILIBook}. While helium burning occurs throughout the profile,
the \Nepa\ reactions are only active towards the end when temperature
and density become high enough.  A nucleosynthesis network containing
583 nuclei is used, which contains nuclei around the nuclear valley of
stability from $A=1$ to $A=100$. This network is sufficient to study
the weak component of the s-process. To characterise the
nucleosynthesis yield uncertainties, final abundances of 4 nuclei are
considered: \nuc{26}{Mg}, \nuc{65}{Cu}, \nuc{70}{Zn}, and
\nuc{87}{Rb}. These are chosen to represent nuclei that are affected
by the \Nepa\ reactions. 

The other three reactions considered here are important in nova
nucleosynthesis. We consider a single zone temperature-density profile
from the hottest hydrogen-burning zone that corresponds to a $1.25
M_{\odot}$ ONe white dwarf accreting material at a rate of $2 \times
10^{-10} M_{\odot}.y^{-1}$. This model is similar to the ``P1'' model
considered by \cite{ILI02} with initial abundances equal to those
presented in their Tab. 2. A nucleosynthesis network containing 146
nuclei from $A=1$ to $A=50$, and linked by 1283 nuclear processes is
used to integrate the system's evolution during a single nova
outburst. Two species of \nuc{26}{Al} are used to represent the
abundances of \nuc{26}{Al} in its ground state, \nuc{26}{Al$^g$}, and
its isomeric state, \nuc{26}{Al$^m$}. It was shown previously by
\cite{ILI02} that some important reactions for this model are
\reaction{18}{F}{p}{$\alpha$}{15}{O},
\reaction{20}{Ne}{p}{$\gamma$}{21}{Na}, and
\reaction{25}{Mg}{p}{$\gamma$}{26}{Al}. The first of these has large
uncertainties arising from ambiguities in the nuclear physics data
\citep[see, for example][]{BEE11}, and strongly affects the final
yields of oxygen and fluorine isotopes. The second reaction considered
is important because at the start of the nova explosion, \nuc{20}{Ne}
comprises 25\% of the material's initial abundance. The reaction
proceeds relatively slowly, thus it strongly affects abundances
throughout the neon-silicon region. The final reaction considered is
important because it directly affects the abundances of magnesium and
\nuc{26}{Al$^g$}. These are important species in nova nucleosynthesis
because they can be measured precisely in pre-solar grains. The
uncertainties for these reactions are adopted from \cite{ILI10b}.

\section{Results and discussion}
\label{sec:results}

\begin{figure}
  \centering
  \includegraphics[width=0.5\textwidth]{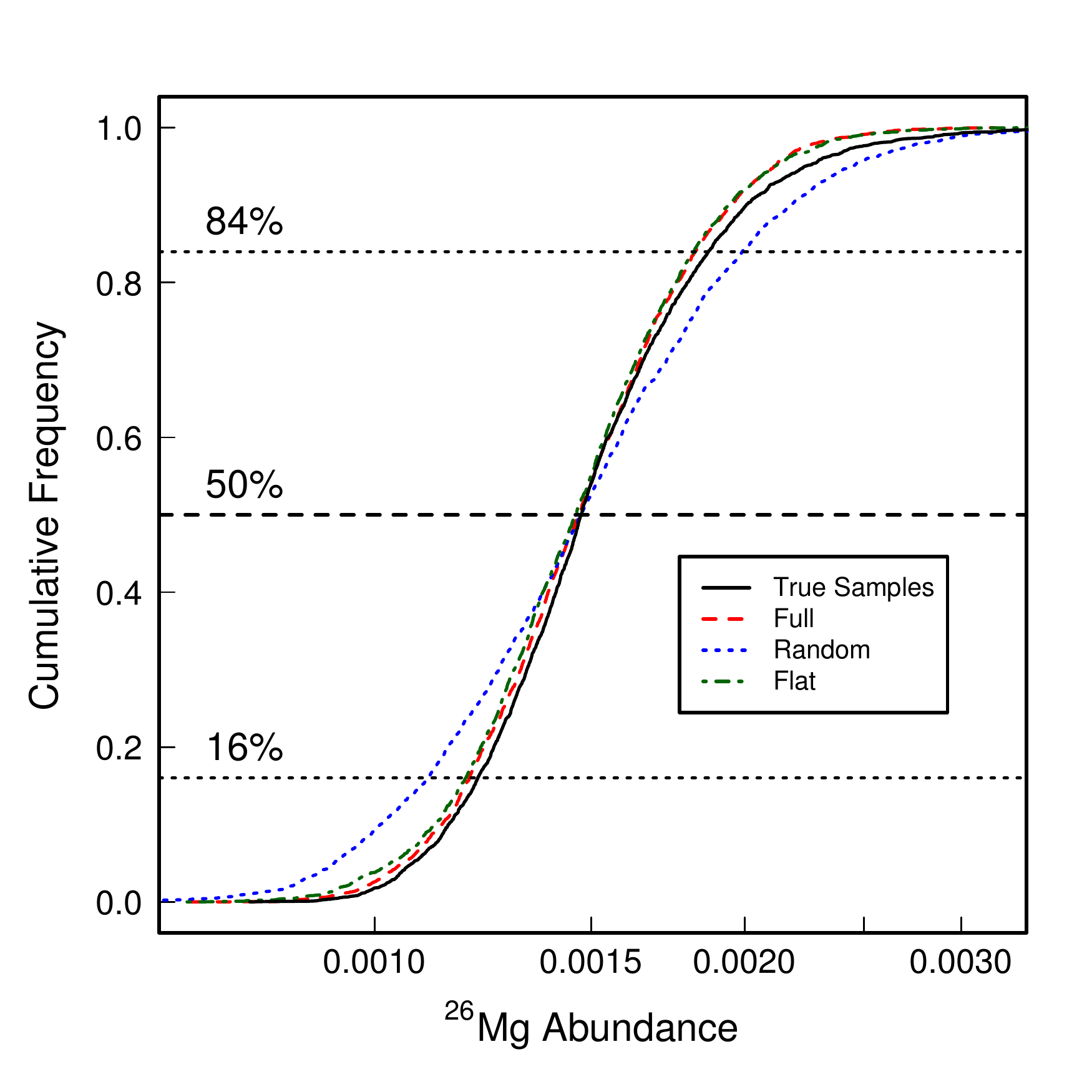}
  \caption{(Colour online) Final \nuc{26}{Mg} abundance cumulative
    density plots the three methods: (i) Input of reaction rate
    samples obtained directly from \texttt{RatesMC} (``Optimum
    Samples''), (ii) full parametrisation of $p$ from
    Eq.~(\ref{eq:tanh}) (``Full''), (iii) randomly sampled reaction
    rates at each temperature (``Random''), and (iv) a constant value
    for $p$ as a function of temperature (``Flat'').}
  \label{fig:Mg26CumDen}
\end{figure}

To illustrate the final distribution of abundances that arise from the
Monte Carlo nucleosynthesis calculations, the final abundances for
\nuc{26}{Mg} obtained from the optimum rate samples, the full and flat
parametrisation schemes, and random sample methods are plotted using
cumulative frequency plots in Fig.~\ref{fig:Mg26CumDen}. A number of
features are immediately obvious. First, the median abundances (i.e.,
the point at which the cumulative distribution crosses the 50\% point)
agree within statistical fluctuation for all methods, as is
expected. This is because, provided proper sampling has been carried
out, the median abundances will correspond to the median rate
regardless of the parametrisation used. Secondly, the random
parametrisation produces a significantly wider distribution than the
results from optimum rate samples. The unphysical, rapidly changing
reaction rates possible when using this method could be responsible
for this, thus producing hard-to-predict nucleosynthesis
results. Indeed, we find that for other reactions such as
\reaction{18}{F}{p}{$\alpha$}{15}{O}, the opposite effect occurs,
i.e., the uncertainties resulting from this method are
\textit{smaller} than those obtained from optimum rate samples. We
conclude, therefore, that this random sampling of rates is not only
unphysical, but also unreliable for calculating accurate
nucleosynthesis yield uncertainties. 
Thirdly, full parametrisation of the
reaction rate samples using the hyperbolic tangent function in
Eq.~(\ref{eq:tanh}) agrees remarkably well with the results from using
the optimum rate samples, albeit with slightly underestimated
uncertainties. 

\begin{figure*}
  \centering
  \includegraphics[width=\textwidth]{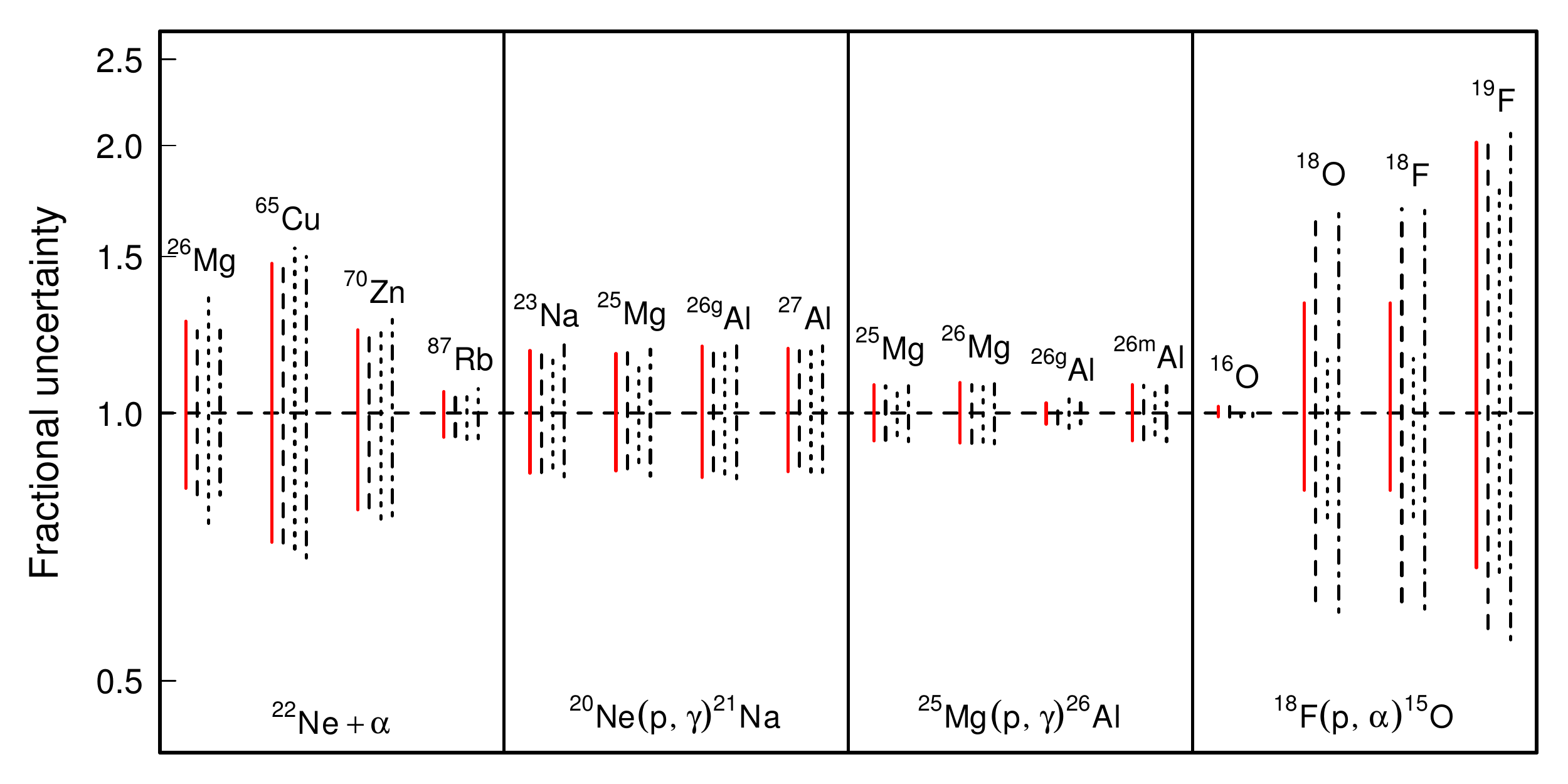}
  \caption{(colour online) The uncertainties computed from 3 of the
    parametrisation schemes investigated here compared to the
    individually sampled reaction rates. The final abundances for four
    nuclei are considered for each reaction under investigation
    here. Each group of vertical bars for each nucleus represents,
    from left to right, the abundances calculated using: (i) The
    "optimum" reaction rate samples; (ii) the full parametrisation; (iii)
    the unphysical, random, case; and (iv) constant values of
    $p(T)=a$.}
  \label{fig:UncertRanges}
\end{figure*}

Perhaps even more remarkable from Fig.~\ref{fig:Mg26CumDen} is that
the uncertainties derived by using flat distributions for $p(T)$ also
agree excellently with the uncertainties derived from using optimum rate
samples, and are barely distinguishable from the full parametrisation
method.  This finding is further enforced in
Fig.~\ref{fig:UncertRanges}, where the uncertainty ranges obtained for
key isotopes arising from varying the four reactions described in
Sec.~\ref{sec:test-cases} are displayed as uncertainty ratios. To
obtain these ratios, uncertainties are first determined from the 16th
and 84th percentiles of their respective cumulative distributions, and
thus represent 1$\sigma$ nucleosynthesis yield uncertainties. These
ranges are then normalised to the median abundance obtained from the
optimum reaction rate samples. Only the ratios for optimum samples, full
parametrisation, random rate samples, and flat parametrisation are
displayed. The ``fixed S and $T_c$'' and ``no offset'' cases are not
displayed for reasons of clarity. For these two cases, the uncertainty
ratios bands are only slightly smaller than those obtained from the
full parametrisation for all cases considered here. This figure
confirms our findings that the full and flat parametrisation schemes
reproduce the uncertainties obtained from optimum rate samples. The
determined uncertainties are all within a few percent of those
obtained from optimum rate samples, and certainly differ by smaller
amounts than the uncertainties, themselves. The random parametrisation
scheme consistently performs poorly, although for some nuclei (i.e.,
those not directly connected to the reaction), the effect is diluted
somewhat.

The only exception to these findings is the
\reaction{18}{F}{p}{$\alpha$}{15}{O} reaction, for which the
uncertainties are too large by roughly a factor of two for \nuc{18}{O}
and \nuc{18}{F} when utilising the parametrisation schemes. The random
parametrisation scheme, contrary to the other reactions, yields
\textit{smaller} nucleosynthesis uncertainties in this case.  The
\reaction{18}{F}{p}{$\alpha$}{15}{O} reaction presents a challenge
because the current uncertainty in the rate is dominated by unknown
factors in the interference effects between a number of
resonances~\citep{ILI10b}. The reaction rate probability density
distribution is, therefore, bi-modal in shape, and hence poorly
represented by a lognormal distribution. This fact is reflected in the
poor Anderson-Darling statistics at most relevant temperatures for
this reaction. The Anderson-Darling statistic represents a measure of
how well the true reaction rate distribution is reproduced by the
lognormal approximation, and is discussed in detail in
\cite{LON10}. In this case, therefore, the $\sigma$ parameter
describing the rate uncertainty is a bad approximation, so using it to
compute reaction rate uncertainties is expected to fail in reproducing
the optimum sample uncertainties. However, for this case, the poor
lognormal shape of the reaction rate distribution already prompts
further experimental investigation. With the addition of new, more
precise nuclear data, the rate will begin to approach a lognormal
distribution more closely, and hence the parametrised Monte Carlo
approach for this reaction will perform better.

Finally, the smallest abundance uncertainties for all nuclei
considered come from using the fixed value for $S$ and $T_c$
parametrisation. This arises from the values adopted for the fixed
parameters, which were chosen to maximise the reaction rate shape
effect discussed before. Although the shape of the rates will be
changing as a function of temperature, the magnitude of the rates
will, on average, be closer to the recommended rate around
T$=0.2$~GK. We can conclude from this that, at least for the present
cases, the shape of the reaction rate has little effect on
nucleosynthesis yields. 

\section{Conclusions}
\label{sec:conclusions}

Reaction rates provided by the recent reaction rate evaluation of
\cite{ILI10a} included temperature dependent lognormal parameters that
were found to accurately reflect the reaction rate probability
densities. However, to date, no investigation has been presented that
determines the most effective way to use these parameters.  Past
research \citep{THE00} showed that nucleosynthesis yields can be
influenced not only by reaction rate magnitudes, but also by their
temperature dependence. An investigation into how best to randomly
sample the reaction rates provided by \cite{ILI10a} is therefore
performed here. The investigation was undertaken for a range of
reactions occurring in core helium burning in massive stars (\Nepa)
and in novae (\neon, \magn, and \flou).


By applying a temperature-dependent rate multiplication factor,
$e^{p(T)\sigma(T)}$, to the reaction rates, the isotopic yields
following post-processing of massive star helium burning and nova
explosion models were investigated for a range of $p(T)$
parametrisation schemes. These choices were compared with the yield
uncertainties obtained when reaction rate samples are obtained
directly from the nuclear physics input and calculated using the
\texttt{RatesMC} reaction rate code from \cite{LON10}.

The most significant finding was that not only can the optimum reaction
rate sample uncertainties be reproduced well by using a hyperbolic
tangent function for $p(T)$, but they are approximated remarkably well
by a flat function: $p(T)=a$. The temperature dependence of the
reaction rates investigated were, therefore, found to be relatively
unimportant to nucleosynthesis yields. This finding is particularly
significant because it means that when performing nucleosynthesis
sensitivity studies, all reaction rates can be sampled
simultaneously. The sensitivity of nucleosynthesis yields to
particular reactions can be characterised by plotting abundance yields
against $p$ (in a similar fashion to Fig. 8 in \cite{PAR08}) and
computing measures of correlation such as the Pearson or Spearman
correlation coefficients. 

For investigating the effect of single reaction rate uncertainties, or
small groups of reactions where correlation coefficients are not
needed, adopting either the full, or flat parametrisation schemes is
recommended. Using the full parametrisation scheme is useful for
investigating the effect that temperature dependence of a reaction
rate has on nucleosynthesis. For rates with large uncertainties, such
as those involving unstable target nuclei, this could still have an
effect, and should always be considered when evaluating reaction rate
uncertainties.

\acknowledgements

This work has been partially supported by the Spanish grants
AYA2010-15685 and by the ESF EUROCORES Program EuroGENESIS through the
MICINN grant EUI2009-04167. I would like to thank Anuj Parikh and
Jordi Jos\'e for their help and recommendations for the manuscript. I
would also like to thank Christian Iliadis for many informative and
enlightening discussions about reaction rate uncertainties.\\


\end{document}